\documentstyle[prl,aps,multicol,psfig,epsfig]{revtex}

\begin{document}
\draft

\title{Unexpected Metallic-like  Behavior of the Resistance
in the Dielectric Spin Density Wave State in
(TMTSF)$_2$PF$_6$}

\author{A.\ V.\ Kornilov$^a$, V.\ M.\ Pudalov$^a$, Y.\ Kitaoka$^b$, K.\
Ishida$^b$, T.\ Mito$^b$, N.\ Tateiwa$^c$,  T.\ C.\ Kobayashi$^c$,
J.\ S.\ Brooks$^d$, and J.\ S.\ Qualls$^d$,
}

\address{
$^{a)}$P.\ N.\ Lebedev Physics Institute, Moscow 119991,  Russia.\\
$^{b)}$Division of Material Physics, School of Engineering
Science, Osaka University, Toyonaka, Osaka 560-8531, Japan\\
$^{c)}$ Research Center for Materials Science at Extreme
Condition, Osaka University, Toyonaka, Osaka 560-8531, Japan \\
$^{d)}$ National High Magnetic Field Laboratory, Florida State
University, Talahassee FL 32310 USA }

\date{\today}
\maketitle

\begin{abstract}
We report unexpected features of the transport in the  dielectric
spin density wave (SDW) phase  of the quasi one-dimensional
compound (TMTSF)$_2$PF$_6$: the resistance exhibits a
maximum and a subsequent strong drop as temperature decreases
below  $\approx 2$\,K. The maximum in $R(T)$
is not caused by depinning or Joule heating of the SDW.
 The characteristic  temperature of the $R(T)$ maximum and the scaling behavior
 of the resistance at different magnetic fields $B$ evidence that the non-monotonic $R(T)$
 dependence has an origin different from
the one known for the quantum Hall effect region of the phase diagram.
We also found that the borderline $T_0(B,P)$ which
divides the field induced SDW region of the $P-B-T$ phase diagram into the
hysteresis and non-hysteresis domains, terminates in the $N=1$
sub-phase; the borderline  has thus no extension to the  SDW $N=0$ phase.

\end{abstract}

\begin{multicols}{2}

The quasi-one-dimensional organic compound (TMTSF)$_2$PF$_6$
undergoes a phase transition from metallic to spin-density wave (SDW)
state as temperature decreases below $T_{\rm SDW}\approx 12$\,K
(at ambient pressure) \cite{gorkov&lebed,gorkov84,ishiguro98,chaikin96}.
As pressure increases,
$T_{\rm SDW}$ decreases and vanishes at $P=6$\,kbar
\cite{gorkov&lebed,gorkov84,ishiguro98,chaikin96}.
Application of the magnetic field $B$ along the least  conducting direction
$z$ (crystal axis $c^*$),
restores the dielectric state; this takes place via  a cascade of the
field induced spin density wave states
(FISDW) accompanied by the sequence of the
quantum Hall effect  (QHE) states  with various
numbers $N$ of filled Landau bands \cite{maki_86,hannahs89,cooper89}. A typical
phase diagram of the FISDW states
for $P=7$\,kbar is illustrated in the inset to Fig.~1.

Earlier \cite{FISDW}, we found that
there is another boundary, $T_0(B)$ which subdivides
the area of the existence of the  FISDW states (at $N \ne 0$)
into the low temperature-  and the
high-temperature domains (see the inset to Fig.~1).
 In the former domain,
the transitions between different FISDW states
($N \Longleftrightarrow N-1$) take place as the first order phase
transitions which manifest experimentally in  hysteretic
variations of observable physical parameters as magnetic
field drives  the system through the transitions.
In the latter domain, the transitions between
different phases are not accompanied with
a hysteresis and are not therefore of the first order.
The above picture is consistent with a novel theoretical model suggested by Lebed
\cite{lebed00} in which the low-temperature and the high-temperature domains have a meaning of the
quantum and semiclassical regions, correspondingly.
In this
novel model,
the nesting vector is predicted to be partially
quantized in the quantum domain and is not quantized
in the semiclassical domain.

We also reported in Ref.~\cite{FISDW} that a qualitative difference between the two domains
exists not only at
the boundaries between the FISDW states, but through the overall area of the  FISDW phase.
We found that the temperature dependence of the resistance $R(T)$
measured along the most conducting direction $x$ (i.e. the crystal axis $a$)
has a maximum in the vicinity of the same $T_0(B)$- line (see  Fig.~1);
this coincidence was observed  for different pressure values from 7 to 14\,kbar.

The maxima in $R_{\rm max}(T)$  were observed earlier \cite{kang92}
in the $N \ne 0$   phases and
have been associated with the onset of the quantum Hall effect \cite{kang92,yakovenko,vuletic}.
Indeed, as $T$ decreases,
the Hall component of the conductivity $\sigma_{xy}$ grows to the quantized value,
 $\sigma_{xy} \rightarrow Ne^2/h$. As a result of the interplay between the
 diagonal and off-diagonal components of conductivity,
 $R_{xx}$ exhibits a maximum, as follows from the equation:
\begin{equation}
R_{xx}={\sigma_{yy}\over\sigma_{xx}\sigma_{yy}+\sigma_{xy}^{2}}
\end{equation}
Within such an explanation (see e.g. Refs.~\cite{kang92,yakovenko} and references therein),
the  $R(T)$-maxima  are associated with the QHE and do exists
in all phases with $N\geq 1$; obviously, such maxima should
be missing
in the dielectric SDW phase with $N=0$.
On the other hand, these maxima in $R(T)$ follow a borderline
$T_0(B)$ which
has a fundamental meaning \cite{lebed00}.
The aim of the current studies  is (i)
to test whether the above explanation holds and $R(T)$
becomes monotonic in the dielectric $N=0$ phase
and (ii) to verify whether or not
the quantum/semiclassical borderline $T_0(B)$ extends to
the dielectric $N=0$
phase as illustrated by the question mark in
the inset to Fig.~1.

We found an unexpected behavior of the resistance
in the  $N=0$ phase: as temperature decreases,
the resistance does not increase monotonically as anticipated for the insulating SDW state
but exhibits a maximum and further falls down by a factor of $\geq 2$.
We found that  the coordinates $T_m(B)$ of the $R(T)$
maximum in the $N=0$ phase
do not fall onto
the $T_0(B)$ borderline extrapolated from  the QHE
region ($N\geq 1$) to the $N=0$ phase.
For example, extrapolated to the same  $P$ and $B$  values,
$T_m$ is typically a factor of 2 lower than $T_0(B)$.
We compared the temperature dependences of the resistance $R(T)$
for $N=0$ and  for the QHE regime at $N\neq 0$
in the vicinity of the $R(T)$-maxima.
The scaling analysis of the two quantities showed that the two effects of maxima in $R(T)$
have different critical exponents and thus the features of the resistance
in the $N=0$ and $N\neq 0$ phases have different origin.
Our results thus demonstrate that the  $T_0(B)$ boundary plotted through the high order
FISDW phases has no extension in the lowest order $N=0$ phase; this is
consistent  with the current theoretical interpretation  \cite{yakovenko}.
On the other hand, the
unexpected $R(T)$
maximum in the purely insulating $N=0$ phase has no explanation
in the frameworks of the current theories.

\vspace{0.1in}
\begin{figure}
\centerline{\psfig{figure=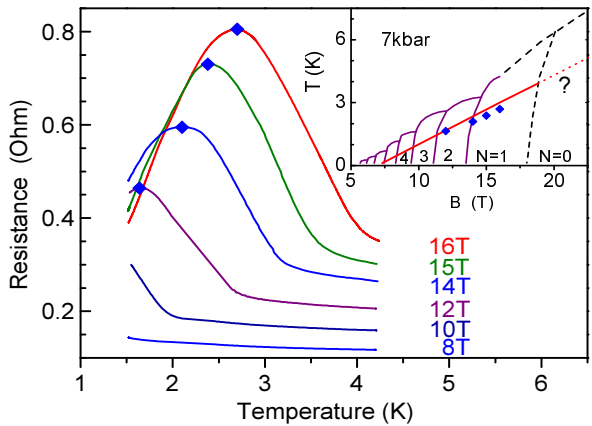,width=220pt,height=150pt}
}
\vspace{0.1in}
\begin{minipage}{3.2in}
\caption{Typical temperature dependence of the
resistance at $P=7$\,kbar, in the FISDW regime for six different values of $B$.
Inset shows the $B-T$-phase diagram: solid curves represent the
results from Ref.~\protect\cite{FISDW},
dashed lines show the phase boundaries anticipated at higher fields according to
 Refs.~\protect\cite{hannahs89,cooper89}.
$N$ values denote the sub-phase number.
The thick straight line
$T_0(B)$ separates the hysteresis and non-hysteresis regimes;
dotted line is its would-be-extrapolation to the $N=0$ phase.
The four diamonds on the main panel and in the inset
depict the  coordinates of the $R(T)$ maxima at various fields.}
\label{fig1}
\end{minipage}
\end{figure}
\vspace{-0.1in}

Measurements were carried out on two samples (of a typical  size $2\times
0.8\times 0.3$\,mm$^3$) grown from a solution by a conventional
electrochemical technique.
We used  either four Ohmic contacts
formed at the $a-b$ plane or eight contacts at two  $a-c^*$ planes;
in all cases $25\mu$m Pt-wires were attached by a graphite paint to the sample
 along the most conducting direction $a$. The sample and a
manganin pressure gauge were inserted  into a Teflon cylinder
placed inside a nonmagnetic 18\,mm o.d. pressure cell
\cite{cylinder_cell} filled with Si-organic pressure transmitting
liquid. The cell was mounted inside the liquid He$^4$, or
He$^3$/He$^4$
mixing chamber,  in a bore of a superconducting magnet. For all
measurements, the magnetic field was applied along $z$.
Sample resistance was measured by four
probe ac technique at 132\,Hz,
with a typical current $1.5\mu$A.
The out-of phase component of the measured
voltage was
negligibly small,
indicating Ohmic contacts to the sample.
The temperature was determined by RuO$_2$ resistance
thermometer. The  temperature was varied slowly, at a rate
$\leq 0.1$\,K/min in order to avoid
deterioration in sample quality. The
changes in the sample resistance were fully reproducible during
the
measurements including temperature sweeps; this
indicated  that the sample quality did not change.
Measurements were done in magnetic fields up to 17.5\,T and  for
temperatures  down to $0.12$\,K.

According to the existing theory \cite{heritier84,lebed85,maki86,yamaji86}
and the known $P-B-T$ phase diagram \cite{phase_diagram,chaikin96},
the insulating $N=0$ phase can be realized in  2 ways:
either at high pressures/high fields
$P>6$\,kbar, $B>18$\,T (as shown in the inset to Fig.~1),
or at low pressures/low fields $P<6$ kbar (as shown in the inset
to Fig.~2). In the latter case,
the magnetic field $B_z$ can be much lower;
it should only be bigger than $0.2$\,T to quench the superconducting state.
In the current  studies we focused on
the low pressure/low field
region in order to avoid possible
influence
of the magnetic breakdown in strong fields
on the
$R(T)$ behavior. Figure 2 demonstrates that the
sample resistance
varies in accordance with the phase diagram:
as temperature decreases, $R$ first decreases
through the metallic phase, then $dR/dT$ changes
sign at the  transition point to the insulating
 SDW phase and $R$ grows by a factor of 40, and,
 eventually, $R$ falls to zero as superconducting state
 sets in. The two squares at the $R(T)$ curve in Fig.~2
 mark the two corresponding transitions.

\begin{figure}
\centerline{\psfig{figure=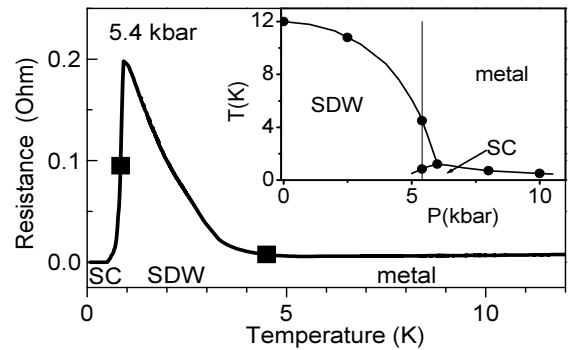,width=210pt,height=130pt}
}
\begin{minipage}{3.2in}
\vspace{0.05in}
\caption{Resistance $R$ vs temperature  measured at
pressure $P=5.4$\,kbar and at $B=0$.
Squares mark the onset of the insulating state at
$T=4.8$\,K and of  the superconducting state at $T=0.9$\,K. The
inset shows $P-T$ phase diagram at $B=0$: dots are the
experimental data, lines are the guide to the eye.}
\label{fig2}
\end{minipage}
\end{figure}
\vspace{-0.1in}

Figure 3 shows the temperature dependence of the resistance in different fields, measured
at pressure 5.4\,kbar.
Starting from high temperatures, $R(T)$ shows a typical metallic behavior. At temperature
$T_{\rm SDW}$, the SDW state (with $N=0$) sets in and the sample resistance starts growing.
Unexpectedly, in this purely insulating state, $R$  exhibits a maximum at a certain temperature
$T_{\rm max} \approx 1$\,K and falls  down by a factor of  2. Similar
behavior of $R(T)$ was observed also at $P=2.5$ and 5.5\,kbar (the results for $P=2.5$\,kbar
are shown  in the inset to Fig.~3).

\begin{figure}
\centerline{\psfig{figure=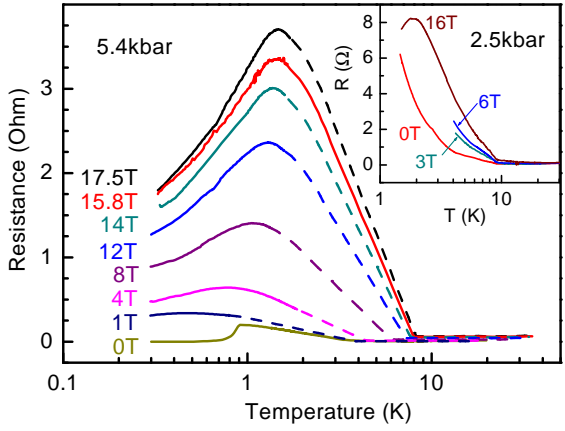,width=210pt,height=160pt}}
\begin{minipage}{3.2in}
\vspace{0.1in}
\caption{Resistance $R$ vs temperature at pressure $P=5.4$\,kbar for different magnetic fields
(indicated on the left side of the curves). Data for $B=0$ and 15.8\,T were measured
overall range of $T$ (solid lines), for other fields data were taken only
at $T<1.7$\,K (solid lines). Dashed
lines depict the anticipated behavior of $R(T)$.
The inset shows similar results for $P=2.5$\,kbar.}
\label{fig3}
\end{minipage}
\end{figure}
\vspace{-0.05in}

The maximum in $R(T)$ is seen for any magnetic field; the temperature of the maximum increases
with field. In order to verify  that the $R(T)$ maxima are not related to depinning or Joule
heating,
we measured the $I-V$ curves in the
vicinity of the $T_{\rm max}$. Figure~4 shows that the differential resistance,  $R=dV/dI$,
on both sides of the maximum, is independent of current up to nearly
4\,$\mu$A;
this result demonstrates that the $R(T)$  maxima measured at $I=1.5$\,$\mu$A
are not  related to the non-ohmic behavior. As  current increases further,
$R$ increases   both, at $T>T_{\rm max}$ and $T<T_{\rm max}$.
This high-current nonlinearity is therefore
likely to be caused by depinning rather than overheating. The former mechanism should
increase the resistance  \cite{vuletic}, whereas
the latter one should cause  $dV/dI$ to decrease  with current
for $T=1.7$\,K, and to increase
 for $T=0.7$\,K (see Fig.~3).

\begin{figure}
\centerline{\psfig{figure=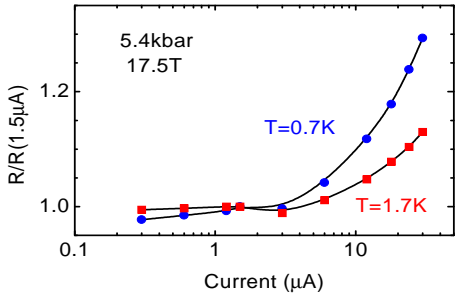,width=170pt,height=110pt}}
\begin{minipage}{3.2in}
\caption{Normalized differential resistance, $R=dV/dI$,
as a function of the current through the sample. Dots and squares
depict the data taken for two temperatures, at both sides of the
$R(T)$-maximum.}
\label{fig4}
\end{minipage}
\end{figure}
\vspace{-0.05in}

The $R(T)$ maxima  seem at first sight similar to the ones which are typical in the FISDW regime
at $N\neq 0$ (shown in Fig.~1). However, the temperatures of the $R(T)$-maximum, $T_{\rm max}$,
in the insulating $N=0$ phase is less than that in the neighboring $N =1$ phase by a
factor of $\gtrsim 2$.
This substantial difference can not be due to the minor difference in pressure
(7\,kbar vs 5.4\,kbar)  because
$T_{\rm max}$ is only weakly pressure dependent
\cite{note}. It follows therefore that the measured $T_{\rm max}$ values in the $N=0$ phase
do not belong to the borderline $T_0(B)$ linearly extrapolated  to the $N=0$ phase
(dashed line in the inset to Fig.~1).
Since we did not observe
other $R(T)$-maxima (i.e., at $T>T_{\rm max}$) in the $N=0$ phase, we conclude
that the semiclassical/quantum borderline, $T_0(B)$, existing throughout the $N>0$ FISDW-phases
does not extend to the insulating $N=0$ phase.

The fundamentally different origin of the $R(T)$-maxima in the FISDW and SDW phases
is demonstrated by the following scaling analysis of the corresponding $R(T)$  data.
In this procedure, we normalized the resistance $R(T)$ (for all curves taken at
different magnetic fields) by its maximum value
$R_{\rm max}=R(T_{\rm max})$
and replaced the temperature $T$ for each curve by the reduced temperature $T/T_{\rm max}$.
Figures~5\,a and b  show the result of such simple scaling for $N=0$ and $N\neq 0$
(i.e. for the SDW and FISDW regimes, correspondingly).

\begin{figure}
\centerline{\psfig{figure=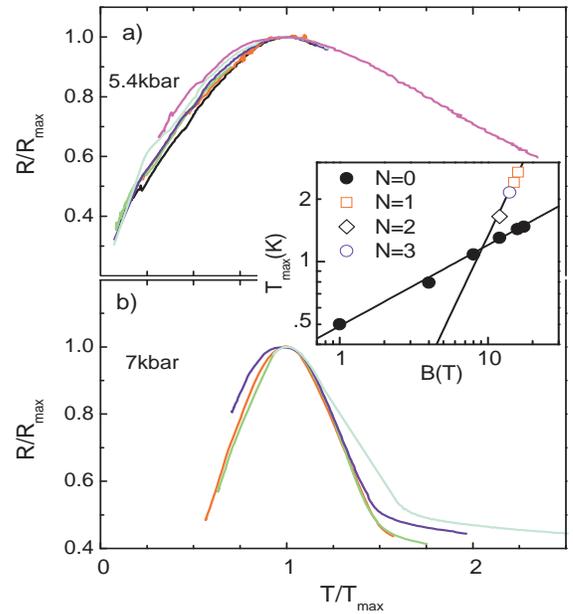,width=210pt,height=230pt}}
\begin{minipage}{3.2in}
\caption{Normalized resistance $R(T)$
vs  normalized temperature
for different fields:
a) for $P=5.4$\,kbar ($N=0$), and b) for $P=7$\,kbar and various $N\neq 0$
(data are calculated from Fig.~3 and Fig.~1, correspondingly).
Inset shows  $T_{\rm max}$ vs magnetic field.}
\label{fig5}
\end{minipage}
\end{figure}
\vspace{-0.1in}

All the data (taken
for different magnetic fields) collapse onto two different universal
dependences $R(T/T_{\rm max})/R(T_{\rm max})$. For $N=0$, all $R(T)$ curves, measured
at magnetic fields from 1 to 17.5\,T  scale excellently thus demonstrating a similar origin.
For $N>0$ (Fig.~5b), the scaling is also good though
there is a minor
systematic departure of the individual curves from the scaling curve, which increases with $N$.
We emphasize that  the two universal scaling curves in Figs.~5\,a and b
are clearly  different; the two scaling exponents
$T_{\rm max}\propto B^\gamma $ are  also different, $\gamma =1.5$ and 0.4, correspondingly
(see the inset to Fig.~5).
This difference demonstrates that the
$R(T)$-maxima  in the SDW phase  and
in the FISDW regime have different underlying mechanisms.

The borderline $T_0(B)$ determined empirically in
Ref.~\cite{FISDW}, separates the regions of the
existence and the absence of the first order
phase transitions.
It was also found that $R(T)$, for different
fields and pressures
exhibits a maximum at nearly the same border.
On the other hand, according to the theory \cite{kang92,yakovenko}
 the maxima  of $R(T)$ in the
 FISDW regime are caused by the developing QHE.
 Our finding that the $T_0(B)$ borderline  has
 no extension to the $N=0$ phase is therefore
in a good agreement with the above interpretation,
because
there is no QHE in the $N=0$ phase.
Our data thus indicates that the QHE develops only in the
`quantum' (hysteretic) domain.
This conclusion suggests  the absence of the QHE in the
high-$N$ phases where $T_0(B)$ vanishes to zero
(e.g., for  $N> 6$ at $P=7$\,kbar,
 as shown in the inset to Fig.~1).

In the purely insulating $N=0$ phase, the QHE
is missing and the appearance of the
 $R(T)$-maxima can not be explained by
 growth of the off-diagonal conductivity.
Therefore, the  origin of the $R(T)$ maxima and of
the strong drop in resistance towards low temperatures
is puzzling. We wish to note an interesting coincidence.
The temperature of the $R(T)$ maxima in the $N=0$
 phase is  $\approx 1.5$\,K in strong magnetic fields ($B=15 - 17$\,T).
On the other hand, the magnetoresistance in the $N=0$ phase exhibits
`rapid oscillations' \cite{RO} whose amplitude vanishes
as  $T$ decreases below approximately the same temperature,  $1.5$\,K.
A symmetry-group analysis \cite{lebed_91}
shows that in (TMTSF)$_2$X compounds there exist two
incommensurate spin density waves. Although the second SDW has
smaller magnitude, it causes `rapid oscillations' and
 may preserve semimetallic properties as
$T\rightarrow 0$.
It is noteworthy, that a low-temperature region of the SDW phase
diagram of (TMTSF)$_2$PF$_6$ was shown to have a non-trivial
origin \cite{takahashi_91,lasjaunias_94} and was suggested
to have an inner structure.
We can not exclude therefore that
the observed by us  metallic-like behaviour of resistivity in
the  $N=0$ phase is somewhat related to inner
subphases of the   $N=0$ SDW phase.

{\it To summarize}, we studied temperature and magnetic field
dependence of the resistance of the quasi-one dimensional
compound (TMTSF)$_2$PF$_6$ both in the QHE ($N\neq 0$) and
SDW ($N=0$) phases. We found that the borderline $T_0(B,P)$
which divides the FISDW region of the $P-B-T$ phase diagram into the
hysteresis and non-hysteresis domains terminates in the $N=1$
sub-phase; the border has no extension to the  SDW $N=0$ phase.
The maxima of $R(T)$ in the FISDW region (which
have been shown to occur approximately at the same borderline)
develop only in the $N\neq 0$ phase; their existence thus correlates
with the existence
of the QHE. This
co-occurrence agrees
with the current theoretical explanation of the $R(T)$-maxima
in the $N\neq 0$ phases.
We found that in the SDW $N=0$ phase which is
considered to be insulating,  the resistance does not
grow infinitely as temperature decreases but exhibits a maximum at
$T_{\rm max} \approx 1-2$\,K and falls to lower
temperatures.
The $R(T)$-maxima in the
`insulating phase' take place in the linear conductance regime
and are not caused by depinning or Joule heating. We found that
the temperature of the $R(T)$-maxima, $T_{\rm max}(B)$,  in the SDW
phase is not a continuation of the border line $T_0(B)$ which
separates the hysteretic and non-hysteretic domains in the $N\neq
0$ regime. A scaling  analysis of the resistance has shown that
the  $R(T)$ maxima in the $N=0$ and $N\neq 0$ phases have different
origin. The unexpected strong drop of the resistance in the SDW $N=0$ phase
at $T<1.5$\,K
 has no explanation within frameworks of the existing
theories.

The work was partially supported by INTAS, RFBR,
NATO, NSF, NWO, Russian Programs `Statistical physics',
`Integration', `The State support of the leading scientific
schools', and COE Research in Grant-in-Aid for Scientific Research, Japan.

\end{multicols}


\vspace{-0.1in}
\begin{references}
\vspace{-0.7in}
\bibitem{gorkov&lebed}L.\ P.\ Gor'kov and A.\ G.\ Lebed, J.\ Phys.
(Paris) Lett. {\bf 45}, L-433 (1984).

\bibitem{gorkov84}L.\ P.\ Gor'kov Sov. Phys. Usp. {\bf 27}, 809 (1984).

\bibitem{ishiguro98}T.\ Ishiguro, K.\ Yamaji and G.\ Saito, Organic
Superconductors (2nd Edition, Springer-Verlag, Heidelberg, 1998).

\bibitem{chaikin96}P.\ M.\ Chaikin J.\ Phys I France {\bf 6}, 1875
(1996).

\bibitem{maki_86}K.\ Maki, Phys. Rev. B {\bf 33}, 4826 (1986).
A.\ Virosztek, L.\ Chen, and K.\ Maki, Phys. Rev. B {\bf 34}, 3371
(1986).

\bibitem{hannahs89}S.\ T.\ Hannahs, J.\ S.\ Brooks, W.\ Kang, L.\ Y.\
Chiang and P.\ M.\ Chaikin Phys. Rev. Lett. {\bf 63}, 1988 (1989)

\bibitem{cooper89}J.\ R.\ Cooper, W.\ Kang, P.\ Auban, G.\ Montambaux
and D.\ J\'{e}rome Phys. Rev. Lett. {\bf 63}, 1984 (1989).

\bibitem{FISDW}A.\ V.\ Kornilov, V.\ M.\ Pudalov, Y.\ Kitaoka, K.\
Ishida, T.\ Mito, J.\ S.\ Brooks, J.\ S.\ Qualls,
J.\ A.\ A.\ J.\ Perenboom, N.\ Tateiwa, T.\ C.\ Kobayashi, cond-mat/0103088.


\bibitem{lebed00}A.\ G.\ Lebed, JETP Lett. {\bf 72}, 141 (2000).

\bibitem{kang92}W.\ Kang, S.\ T.\ Hannahs, L.\ Y.\ Chiang, R.\ Upasani,
P.\ M.\ Chaikin, Phys. Rev. B {\bf 45}, 13566 (1992).

\bibitem{yakovenko}V.\ M.\ Yakovenko, H.-S.\ Goan, Phys. Rev. B
{\bf 58}, 10648 (1998).

\bibitem{vuletic}T.\ Vuleti\'{c} C.\ Pasquier, P.\ Auban-Senzier,
S.\ Tomi\'{c}, D.\ J\'{e}rome, K.\ Maki, and K.\ Bechgaard,
Eur. Phys. J. B {\bf 21}, 53 (2001).

\bibitem{cylinder_cell}A.\ V.\ Kornilov, V.\ A.\ Sukhoparov, V.\ M.\
Pudalov, {\it High Pressure Science and Technology}, ed. W.\
Trzeciakowski, World Scientific, Singapore, 63 (1996).

\bibitem{heritier84}M.\ Heritier, G.\ Montambaux and P.\ Lederer, J.\
Phys. (Paris) Lett. {\bf 45}, L-943 (1984).


\bibitem{lebed85}A.\ G.\ Lebed, Sov. Phys.: JETP, {\bf 62}, 595 (1985).

\bibitem{maki86}K.\ Maki, Phys. Rev. B {\bf 33}, 4826 (1986).
\bibitem{yamaji86}K.\ Yamaji, Synth. Met. {\bf 13}, 29 (1986).

\bibitem{phase_diagram}W.\ Kang, S.\ T.\ Hannahs, P.\ M.\ Chaikin,
Phys. Rev. Lett. {\bf 70}, 3091 (1993).

\bibitem{note} Figure~3 shows that the $T_{\rm max}$-values measured  at $P=2.5$ and 5.4\,kbar differ
from each other less than by 20\%.

\bibitem{RO}J.\ P.\ Ulmet, L.\ Bachere, and S.\ Askenazy, Sol. St. Commun. {\bf 58},
753 (1986). S.\ Uji, J.\ S.\ Brooks, M.\ Chaparala, S.\ Takasaki, J.\ Yamada, and H.\ Anzai,
Phys. Rev. B {\bf 55}, 12446 (1997).

\bibitem{lebed_91}A.\ G.\ Lebed, Physica Scripta, {\bf 39}, 386
(1991).

\bibitem{takahashi_91}T.\ Takahashi, T.\ Harada, Y.\ Kobajashi,
K.\ Kanoda, K.\ Suzuki, K.\ Murata, and G.\ Saito, Synth. Met.,
{\bf 41-43}, 3985 (1991).

\bibitem{lasjaunias_94}J.\ C.\ Lasjaunias, K.\ Biljakovi\'{c}, F.\
Nad', P.\ Monceau, K.\ Bechgaard, Phys. Rev. Lett. {\bf 72}, 1283
(1994).


\end{references}
\end{document}